# Estimation of Motion Parameters for Ultrasound Images Using Motion Blur Invariants

Barmak Honarvar Shakibaei
Cranfield University, School of Aerospace, Transport and Manufacturing, Bedfordshire MK43 0AL, UK
+44 (0)1234-75-4128
barmak@cranfield.ac.uk

Yifan Zhao
Cranfield University, School of Aerospace, Transport and Manufacturing, Bedfordshire MK43 0AL, UK
+44 (0)1234-75-4729
yifan.zhao@cranfield.ac.uk

John Ahmet Erkoyuncu
Cranfield University, School of Aerospace, Transport and Manufacturing, Bedfordshire MK43 0AL, UK
+44 (0)1234-75-4717
j.a.erkoyuncu@cranfield.ac.uk

## ABSTRACT
The quality of fetal ultrasound images is significantly affected by motion blur while the imaging system requires low motion quality in order to capture accurate data. This can be achieved with a mathematical model of motion blur in time or frequency domain. We propose a new model of linear motion blur in both frequency and moment domain to analyse the invariant features of blur convolution for ultrasound images. Moreover, the model also helps to provide an estimation of motion parameters for blur length and angle. These outcomes might imply great potential of this invariant method in ultrasound imaging application.

## 1. INTRODUCTION

Improving the quality of ultrasound images is a difficult task because their characteristic feature is speckle. Speckle arises from signal interference caused by tissue microinhomogeneities (tissue cells, capillaries, blood cells, etc). This coherent summation of back scattered signals forms a spatial distribution of speckle that is specific to the density and distribution of the scatterers and thus to the nature of the tissue [1]. Image blurring is frequently an issue that affects the performance of an identification system. Blur may arise due to diverse sources like atmospheric turbulence, defocused lens, optical abnormality, and spatial and temporal sensor assimilation. Two common types of blurs are motion blur and defocus blur. Motion blur can occur when a tissue structure moves temporarily out of the imaging plane (due to its own motion, the motion of the transducer, or that of the patient) while the defocus blur is caused by the inaccurate focal length adjustment at the time of image acquisition. Blurring induces the degradation of sharp features of image like edges, specifically for ultrasound images where the encoded information is easily lost due to blur. The point spread function (PSF) of an imaging system introduces some levels of blurring in the captured images. Mostly the PSF is modeled as a Gaussian distribution which is widely applicable in imaging devices [2]. In real applications, images contain various artifacts such as geometrical and convolutional degradations. Image analysis systems should be able to operate also in these non-ideal conditions. There has been a vast amount of research in this field of invariant pattern for object recognition [3]. However, the invariant recognition of objects degraded by blur is a much less studied topic in medical imaging research. This degradation process can be modeled as a linear shift-invariant system in which the relation between an ideal image $f(x,y)$ and an observed image $g(x,y)$ is given by

$$g(x,y) = f(x,y) * h(x,y) + n(x,y), \quad (1)$$

where $(x,y)$ represents a 2-D spatial pixel coordinate, $h(x,y)$ is the PSF of the system, $n(x,y)$ is additive noise, and $*$ denotes 2-D convolution. The PSF represents blur while other degradations are captured by the noise term $n(x,y)$. This blurring effect causes a significant reduction in the sharpness of compound images, especially in ultrasound motion sequences. Generally, the more frames used for compounding, the greater the improvement in image quality and the greater potential for motion blurring. This results in a trade-off between improving image quality and minimizing motion blurring.

Invariant properties have been traditionally used to characterize the spatial distribution of patterns. Besides studies that aimed to directly estimate the speckle distribution, the use of general invariant analysis methods had some success in ultrasound image and their segmentation (see [4] for a survey). However, motion analysis needs relatively large windows to perform feature estimation; this leads to a lack of precision, especially at tissue boundaries. Other ultrasound image characteristics can also make feature extraction more problematic. Because of the ultrasound beam form, the size of the motion increases according to the distance from the ultrasound probe. Moreover, for circular probes, the ultrasound beam directions radiate from the probe center because of its geometry [5]. These radial directions have a direct impact on the motion orientations, which are different depending on the motion direction across the image. In order to obviate velocity- and blur-estimation, Levin *et al.* have proposed motion invariant image capture for moving subjects [6]. In order to demonstrate the concept, prototype cameras were developed based on whole camera rotation and sensor shifting using custom hardware [6]. In both cases, image stabilization hardware has been mentioned as a preferred implementation of motion invariant image capture.

Here, we focus on establishing a novel methodology to analyze the motion blur invariants in both frequency and moment domains. Then we estimate the angle and length of blur as motion parameters in ultrasound images using PSF spectrum of motion blur.

## 2. MOTION BLUR FORMULATION

The blurring of images is modeled in (1) as the convolution of an ideal image with a 2D PSF. It is worth noticing that PSFs are not a function of the spatial location under consideration, i.e., they are spatially invariant. In most cases the blurring of images is a spatially continuous process. Since identification and restoration methods are always based on discrete images, we present the blur model in its continuous form, followed by its discrete (sampled) counterparts. A relative motion between the sensor and the scene during the exposure interval causes the light field image to be motion blurred. Many types of motion blur can be distinguished all of which are due to relative motion between the recording device and the scene. This can be in the form of a translation, a rotation, a sudden change of scale, or some combinations of these. With an introduction to the terms blur angle $\theta$ and blur length $L = v_0 t$ (where $v_0$ is the velocity of camera movement and $t$ is the exposure time), the motion blur PSF is given by:

$$h(x,y) = \frac{1}{L} \int_{-L/2}^{L/2} \delta(x - t\cos\theta)\delta(y - t\sin\theta) dt, \quad (2)$$

where $\delta(.)$ is the Dirac function. The discrete form of (2) is not easily captured in a close form expression in general. In case of linear horizontal motion, the blur angle is zero ($\theta = 0°$) and $h(x,y) = \frac{1}{L}\delta(y)$ for $|x| \leq L/2$. For linear vertical motion, the blur angle is 90° and $h(x,y) = \frac{1}{L}\delta(x)$ for $|y| \leq L/2$. We use an approximated discrete form of (2) by applying the relation of discrete delta function to rectangle function, $\Pi(.)$, as follows:

$$\begin{aligned} h(n,m) &= \frac{1}{L}\sum_{k=-\infty}^{+\infty} \delta(n - k\cos\theta)\delta(m - k\sin\theta) \\ &= \frac{1}{L}\Pi(m\cos\theta - n\sin\theta). \end{aligned} \quad (3)$$

It is possible to show that the behaviour of the equations (2) and (3) is almost same. Figure 1(a)-(b) show a sharp fetus image and its degraded result caused by a linear motion, respectively. When the gradient of such a blurred image is transformed into the frequency domain, a series of bright-dark parallel stripes are contained, as shown in figure 1(c). Moreover, figure 1(d) illustrates the clues for PSF identification from the periodical distribution of dark stripes. It is clear that the blur angle $\theta$ is equivalent to the angle between the parallel dark stripes and the image vertical axis, while the blur length corresponds to the distance between neighboring dark stripes [7]. Therefore, the blur parameters can be determined by calculating the numeric characteristics of these dark stripes.

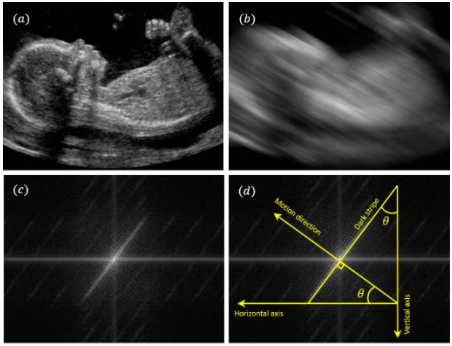

**Figure 1. (a) Fetus ultrasound image, (b) motion blurred image, (c) frequency spectrum image of (b), (d) the relationship between the blur angle $\theta$ and motion direction in (c).**

Figure 2 shows the density graph of the discrete PSF for $L = 60$ and different values of $\theta$ on the left side. On the right side of the same figure, the fetus ultrasound image shown in figure 1(a) is simulated with a fixed length of motion blur and the same various motion angles to illustrate the visualization of motion blur.

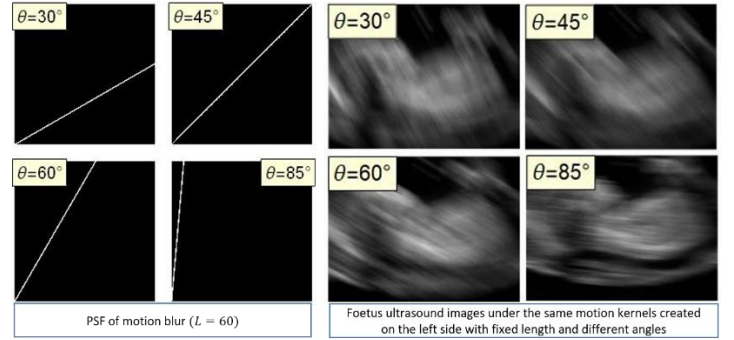

**Figure 2. Understanding of motion blur. Left: PSF of motion blur in spatial domain for $L = 60$ and different angles ($\theta = 30°, 45°, 60°, 85°$) and right: Fetus ultrasound images with fixed length and different angles of motion.**

## 3. MOTION BLUR INVARIANT

In this section, we establish a new frequency and moment blur invariant based on a linear motion PSF for degraded images. Since an imaging system can be modeled as a 2D convolution in (1), it is possible to transform this equation to the Fourier or moment domains. For frequency analysis, we consider the imaging system in the presence and absence of noise, respectively. For moment domain analysis, we only derive the invariant properties of ultrasound images in the absence of noise.

### 3.1 Frequency Domain Invariant

**Effect of Noise.** In the presence of noise, the degradation model in (1) can be expressed in the Fourier domain as:

$$G(u,v) = F(u,v)H(u,v) + N(u,v), \quad (4)$$

where $G(u,v)$, $F(u,v)$, $H(u,v)$ and $N(u,v)$ are the frequency responses of the observed image, original image, PSF, and noise, respectively. The Wiener deconvolution method has widespread use in image deconvolution applications, as the frequency spectrum of most visual images is fairly well behaved and may be estimated easily. Here, the target is to find $\lambda(x,y)$ in the way that $\hat{f}(x,y)$ can be approximated as a convolution, that is, $\lambda(x,y) * g(x,y)$, to minimize the mean square error, where $\hat{f}(x,y)$ is an estimation of $f(x,y)$. The Wiener deconvolution filter provides such a $\lambda(x,y)$. The filter is described in the frequency domain [8, 9]:

$$\Lambda(u,v) = \frac{\overline{H(u,v)}S(u,v)}{|H(u,v)|^2 S(u,v) + N(u,v)}, \quad (5)$$

where $S(u,v)$ the mean power spectral density ($S(u,v) = E\{|F(u,v)|^2\}$) of the original image, $f(x,y)$ and the vinculum denotes complex conjugation. Using this technique to find the best reconstruction of a noisy image can be compared with other algorithms such as Gaussian filtering.

**Absence of Noise.** If noise is neglected, Eq. (4) can be reduced to a simple product of F(u, v) and H(u, v). The Fourier transform of (2) can be written as the following sinc function:

$$H(u,v) = sinc\left(\frac{L(u\cos\theta + v\sin\theta)}{2\pi}\right). \quad (6)$$

Figure 3 shows the Fourier transform of the PSF of motion blur with different values of blur lengths and angles. The figure illustrates that the blur is effectively a low-pass filtering operation and has spectral zeros along characteristic lines.

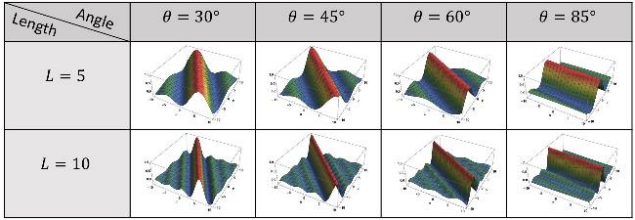

**Figure 3. PSF of motion blur in the Fourier domain for different lengths ($L = 5, 10$) and angles ($\theta = 30°, 45°, 60°, 85°$).**

The imperfections in the image formation process are modeled as passive operations on the data, i.e., no "energy" is absorbed or generated. Consequently, for spatially continuous blurs the PSF is constrained to satisfy the global energy preserving $\int_{-\infty}^{\infty}\int_{-\infty}^{\infty} h(x,y)dxdy = 1$ which also supports the spectral condition of $H(0,0) = 1$.

By using the reduced version of (4) in the absence of noise, we get

$$G(u,v) = F(u,v)sinc\left(\frac{L(u\cos\theta + v\sin\theta)}{2\pi}\right). \quad (7)$$

To find the motion blur parameters (length and angle), we set both frequencies $(u, v)$ to $(0,1)$ and $(1,0)$, respectively. By defining $a = sinc^{-1}[G(0,1)/F(0,1)]$ and $b = sinc^{-1}[G(1,0)/F(1,0)]$, we derive $\theta = \tan^{-1}(a/b)$ and $L = 2\pi a \csc\theta$. Finally, by substituting the obtained angle and length of motion blur in terms of low frequencies, $(u, v) \in \{0,1\}$, in (7), we can derive a frequency invariant scheme for motion blur as:

$$\xi(u,v) = \frac{G(u,v)}{F(u,v)} = sinc[a(u\cot\theta + v)] = sinc(av + bu). \quad (8)$$

It means that there is a relationship between the original and degraded images' spectrum ($F(u,v)$ and $G(u,v)$) with their low frequency components ($F(0,1)$, $F(1,0)$, $G(0,1)$ and $G(1,0)$). Eq. (8) shows the proposed blur invariant features in Fourier domain - called $\xi(u,v)$ - for all range of frequencies which is independent of the motion blur kernel parameters. In Section 4, we show some of these invariants.

## 3.2 Motion Blur Invariant in Moment Domain

As specified in the introduction, the motion blur analysis on an ultrasound image should be performed using some descriptors that are invariant to convolution, rotation and scaling because the speckle size is directly related to the ultrasound beam diameter which is not constant over the image. In this paper, we use a set of geometric moment invariant reported in [10]. Ignoring the additive noise $n(x,y)$ in (1), the following relation between the observed image moments and the original and PSF moments exists [10]

$$m_{pq}^{(g)} = \sum_{k=0}^{p}\sum_{l=0}^{q}\binom{p}{k}\binom{q}{l}m_{kl}^{(h)}m_{p-k,q-l}^{(f)}, \quad (9)$$

where $m_{pq}^{(g)}$, $m_{pq}^{(f)}$ and $m_{pq}^{(h)}$ are the two-dimensional $(p+q)^{th}$ order geometric moments of the observed image, original image, and PSF respectively. The two-dimensional $(p+q)^{th}$ order geometric moments of the original image is defined by

$$m_{pq}^{(f)} = \int_{R^2} f(x,y)x^p y^q dxdy. \quad (10)$$

The geometric moments of the PSF can be calculated from the motion function (using sifting property of Dirac delta function) with Eq. (2):

$$m_{pq}^{(h)} = \begin{cases} \frac{(L/2)^{p+q}(\cos\theta)^p(\sin\theta)^q}{p+q+1}, & p+q = \text{even} \\ 0, & p+q = \text{odd}. \end{cases} \quad (11)$$

Substituting (11) in (9) and expanding the observed image moments in terms of the original image moments, it is clear that the zeroth and first invariant moments could be found directly ($m_{00}^{(g)} = m_{00}^{(f)}, m_{01}^{(g)} = m_{01}^{(f)}, m_{10}^{(g)} = m_{10}^{(f)}$). The second orders invariant can be obtained as follows:

$$\begin{aligned} m_{11}^{(g)} &= (L^2/24)\sin2\theta \ m_{00}^{(f)} + m_{11}^{(f)} \\ m_{20}^{(g)} &= (L^2/12)\cos^2\theta \ m_{00}^{(f)} + m_{20}^{(f)} \\ m_{02}^{(g)} &= (L^2/12)\sin^2\theta \ m_{00}^{(f)} + m_{02}^{(f)} \end{aligned} \quad (12)$$

From the last two equations in (12), we can find the direction of motion blur in terms of the second order moments as follows:

$$\theta = \tan^{-1}\left(\frac{m_{02}^{(g)} - m_{02}^{(f)}}{m_{20}^{(g)} - m_{20}^{(f)}}\right)^{1/2}. \quad (13)$$

Finally, by substituting (13) in the second equation of (12), the length of the motion blur could be obtained as:

$$L = 2\sqrt{3}\left(\frac{m_{20}^{(g)} + m_{02}^{(g)} - m_{20}^{(f)} - m_{02}^{(f)}}{m_{00}^{(f)}}\right)^{1/2}. \quad (14)$$

Eqs. (9), (13) and (14) show that the moment invariants are a linear combination of their original moments, thus they maintain the capacity for feature analysis. Moreover, these derivations confirm that our proposed invariant scheme is matched with the ordinary moment invariants with respect to blur (convolution). In the result section, we evaluate these invariants of different orders ($M_{(p+q)}$) for degraded ultrasound images by motion blur.

## 3.3 Motion Blur Parameter Estimation in Ultrasound Images

As a matter of fact, PSF is decided by two motion parameters: the direction and the angle of motion, whose values are often unavailable due to the intrinsic nature of ultrasound motions and speckle. An improved solution is the blind image deblurring techniques [11], which extracts motion parameters from the blurred image and then restores the true appearance with the estimated PSF.

**Blur Angle Estimation.** As mentioned previously, the blur angle can be obtained by measuring the direction of the approximately linear dark stripes in frequency spectrum. In order to heighten the estimation accuracy, a bilateral piece-wise estimation strategy is proposed based on the principle of error suppress. First of all, two approximately linear edges on both sides of the central bright stripe are extracted by means of classical edge detection algorithm. Then the identified edges are divided into several overlapping small segments, and the angle between each segment and vertical axis is individually estimated, constituting a series of estimation values of motion direction. With these angles, the most appropriate representative of the true blur angle can be finally calculated through an effective information fusion method.

**Blur Length Estimation.** Once finding the blur angle with the described above algorithm, we can take the discrete Fourier transform of the corresponding PSF and it would be a discrete version of (6) as

$$H(u,v) = sinc\left[\frac{L}{2\pi}\left(\frac{u\cos\theta}{M} + \frac{v\sin\theta}{N}\right)\right], \quad (15)$$

where $M$ and $N$ are the size of image. Assuming $\omega = (u\cos\theta)/M + (u\sin\theta)/N$, one way to find the location of dark strips, $H(\omega) = 0$ should be solved, which would generate the following formula:

$$\frac{u\cos\theta}{M} + \frac{v\sin\theta}{N} = \pm\frac{2k\pi}{L}. \quad ; \quad k = 1,2,\dots \quad (16)$$

Notice that when the image is square ($M = N$), the above formula can be reduced to $u\cos\theta + v\sin\theta = \pm 2kN\pi/L$. From (16), we conclude that $L$ determines the position of dark strips in frequency domain. In other words, the length of motion can also be expressed in terms of $D$, the distance between two consecutive zeros and the image size ($M \times N$) as $L = (M \times N)/D$.

## 4. EXPERIMENTS

Three numerical experiments are conducted in order to prove the validity and the efficiency of the proposed methods. The first and the third experiments have been performed using two sets of videos (five slow and five fast scans) without any visual feedback in a trajectory (axially from head to toe or toe to head followed by moving the probe in the opposite direction after placing it in a perpendicular orientation) based on scans of a fetal phantom (SPACEFAN-ST, Kyoto Kagaku) by a convex transducer probe with a Telemed MicrUs Scanner (Telemed Ultrasound Medical Systems, Lithuania). In the second experiment, we used an 11 weeks' gestation fetal ultrasound image.

### 4.1 First Experiment

Table 1 shows the slow and fast scan with different motion blur levels. The blur invariants shown in (8) are denoted as $\xi(u,v)$ where the frequencies, $u$ and $v$, are varied in random ranges. In each row of this table, the results of the amplitude and phase of slow/fast scan invariants are shown. It can be observed that their respective values change slightly for slow/fast motions.

### 4.2 Second Experiment

An original 11 weeks' gestation fetal ultrasound image was blurred by four different motion blur of the direction 30°, 45°, 60° and 85°. The length of blurs for the corresponding directions were $L = 20$, $L = 40$, $L = 30$ and $L = 50$, respectively. For these five images we calculated blur moment invariants based on (9), (13) and (14) from zeroth order to fourth order (see Table 2).

| Invariant | 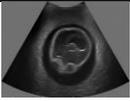 | 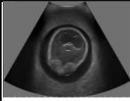 | 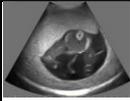 | 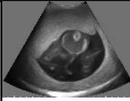 |
|---|---|---|---|---|
| | Slow scan | Fast scan | Slow scan | Fast scan |
| $\|\xi(1,2)\|$ | 107.954 | 109.011 | 122.548 | 120.123 |
| $\angle\xi(1,2)$ | -0.9815 | -0.9791 | -0.4618 | -0.4108 |
| $\|\xi(2,3)\|$ | 97.748 | 95.125 | 136.782 | 138.944 |
| $\angle\xi(2,3)$ | -3.0246 | -3.1108 | 2.9232 | 3.1005 |
| $\|\xi(5,5)\|$ | 170.046 | 167.191 | 116.087 | 112.116 |
| $\angle\xi(5,5)$ | 2.8055 | 2.8397 | 1.9931 | 1.7895 |
| $\|\xi(7,11)\|$ | 116.313 | 119.216 | 139.203 | 138.019 |
| $\angle\xi(7,11)$ | 0.1167 | 0.1238 | 1.5041 | 1.5218 |
| $\|\xi(4,25)\|$ | 149.573 | 146.911 | 192.149 | 195.333 |
| $\angle\xi(4,25)$ | -1.8841 | -1.8519 | -0.5029 | -0.4991 |

**Table 1.** The values of the frequency invariants including amplitude ($|\xi(u,v)|$) and phase ($\angle\xi(u,v)$ in radian) with different values of $(u,v)$ for slow and fast scan ultrasound images followed by motion model.

One can see from Table 2 how important is to understand the theoretical properties of the moment invariants under various level of motion blur. As we discussed in subsection 3.2, the zeroth and the first invariant moments are the same (row $M_0$ and $M_1$ in table). On the other hand, all $M_2$, $M_3$ and $M_4$ invariants provide a perfect stability.

| Invariant | 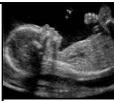 | 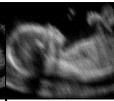 | 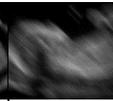 | 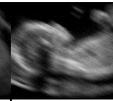 | 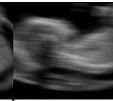 |
|---|---|---|---|---|---|
| | Original | $\theta = 30°$, $L = 20$ | $\theta = 45°$, $L = 40$ | $\theta = 60°$, $L = 30$ | $\theta = 85°$, $L = 50$ |
| $M_0$ | 4.5264 | 4.5264 | 4.5264 | 4.5264 | 4.5264 |
| $M_1$ | 3.1773 | 3.1773 | 3.1773 | 3.1773 | 3.1773 |
| $M_2$ | 1.0254 | 1.1105 | 0.9992 | 1.0138 | 0.9987 |
| $M_3$ | 0.3621 | 0.3549 | 0.3618 | 0.3657 | 0.3592 |
| $M_4$ | 2.9244 | 2.9637 | 3.0025 | 2.8964 | 2.9150 |

**Table 2:** The values of the geometric moment invariants with different motion blur parameters of angle/length. $M_r$ shows the invariant value of the moment order $r$, as discussed in subsection 3.2.

### 4.3 Third Experiment

Finally, three sets of fetal phantom images are used to estimate motion blur parameters based on subsection 3.3. Moreover, a comparative analysis is performed with the classical Cepstrum domain algorithm [12]. Table 3 presents the comparative analysis of the proposed work with this classical method evaluated in terms of BRISQUE and SSIM scores [13, 14]. It can be observed that there is significant noticeable artifacts at the borders of the restored image using [12]. The proposed method performs competitively when compared to the existing methods. It is worth noting that both methods are quite accurate in terms of blur angle and length estimation while the proposed method provides better image quality with respect to image scores.

| Blurred image | Deblurred image | |
|---|---|---|
| | Cepstrum domain [12] | Frequency domain |
| 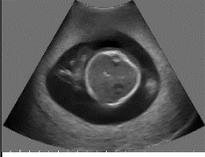 | 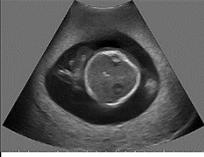 | 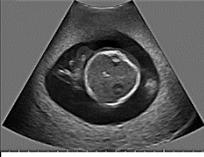 |
| Fast scan | $\hat{L} = 16$ , $\hat{\theta} = 22°$ | $\hat{L} = 15$ , $\hat{\theta} = 23°$ |
| BRISQUE: 40.7475 | 37.8351 | 29.1075 |
| SSIM: NA | 0.9273 | 0.9671 |
| 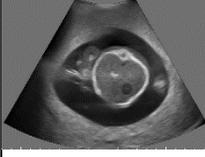 | 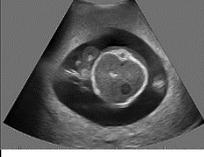 | 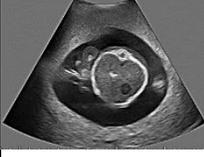 |
| First slow scan | $\hat{L} = 7$ , $\hat{\theta} = 76°$ | $\hat{L} = 7$ , $\hat{\theta} = 75°$ |
| BRISQUE: 42.3254 | 34.5529 | 25.6311 |
| SSIM: NA | 0.9466 | 0.9772 |
| 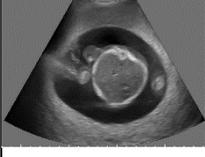 | 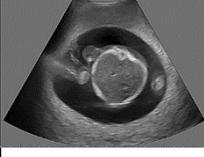 | 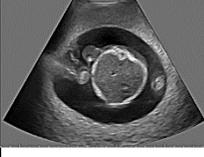 |
| Second slow scan | $\hat{L} = 5$ , $\hat{\theta} = 33°$ | $\hat{L} = 5$ , $\hat{\theta} = 33°$ |
| BRISQUE: 41.2873 | 35.3568 | 24.3949 |
| SSIM: NA | 0.9355 | 0.9731 |

**Table 3**: Image deblurring in Cepstrum (method [12]) and frequency domains (proposed scheme) using estimated motion algorithm and their corresponding BRISQUE/SSIM scores.

## 5. CONCLUSION

In this paper, we have proposed a suitable model of fetal ultrasound imaging systems with respect to their motion blur phenomenon. The idea of invariant features of fast and slow scan of ultrasound images is developed in both frequency and moment domains. We studied the PSF behaviour of blurry ultrasound images in time-domain, moment domain, matrix form and frequency domain. An estimation algorithm of PSF in terms of blur angle and blur length is also proposed. Using the obtained PSF information, restoration of the motion blurred ultrasound images is performed in frequency domain. Experiments demonstrate that better perceptual quality was obtained in frequency domain. A comparative analysis of deblurring results obtained using the cepstrum and frequency domains are conducted using BRISQUE and SSIM scores. It has been observed that using frequency domain, results are robust to the variation in the estimated PSF parameters.

## 6. ACKNOWLEDGMENTS

This work was supported by the UK EPSRC GCRF Grant: Distributed Intelligent Ultrasound Imaging System for Secure in-community Diagnostics (SecureUltrasound) (Grant number EP/R013950/1). The work of Barmak Honarvar Shakibaei Asli was supported by the Czech Science Foundation under Grant 18-26018Y.